\documentstyle[prl,aps,graphicx]{revtex}
\newcommand{\beqa}{\begin{eqnarray}}
\newcommand{\eeqa}{\end{eqnarray}}
\def\build#1_#2{\mathrel{\mathop{\kern 0pt#1}\limits_{#2}}}

\title{A mode-coupling theory for the pasty rheology of soft glassy materials}

\author{ P. H\'ebraud~\footnote{Present adress : Laboratoire PCC, Institut Curie, 11 rue Vauquelin, 75005 Paris FRANCE {\it e-mail : pascal.hebraud@curie.fr}} and F. Lequeux~\footnote{ Present adress :  Laboratoire de physicochimie macromol\'eculaire, ESPCI, 10 rue Vauquelin, 75231 Paris FRANCE {\it e-mail : francois.lequeux@espci.fr}}}

\address{Laboratoire de Dynamique des Fluides Complexes, 3 rue de l'Universit\'e, 67084 Strasbourg FRANCE}
\begin{document}
\draft
\maketitle

\vspace{0.5truecm}
\begin{abstract}
In this paper, we introduce a simple mode-coupling model for concentrated suspensions under flow. This model exhibits a jamming transition, and stress {\it vs} shear rate relations which are very similar to experimental results. Namely newtonian regime or yield stress are followed by a slow variation of the stress for higher shear rates, and by an apparent newtonian regime for very large shear rates. Another striking result is that under oscillating strain, even in the jammed state, the system exhibits a relaxation time which depends on the strain amplitude.
\end{abstract}


\vspace{.5truecm}

\indent	We present a model which aims at describing the mechanical behavior of very concentrated suspensions of soft particles. Such systems, like concentrated emulsions, colloidal suspensions near random close packing, concentrated suspensions of small particles of gel, and concentrated colloids electrostatically charged, are known to exhibit peculiar paste-like behavior for high enough concentration of the dispersed phase : they have a yield stress and thus very non-linear mechanical properties. They all share the property of being made of a collection of soft objects.\\
 Beyond a critical value of their concentration, the particules become packed together, and they experience a jamming transition. These systems have recently been named Soft Glassy Materials.\\
\indent At the macroscopic scale, the jamming transition corresponds to the fact that, below a critical concentration, the system is fluid, and the stress vanishes at rest ; but, above the critical concentration, stress can be stored in the system --- after flow cessation for instance ---. Obviously, because these systems exhibit a yield stress, their mechanical behavior under the action of flow is very non-linear. More precisely, some recent model systems and simulations exhibit a power-law relation between stress and shear rate, with quite small exponents --- around $.1$ --- for small shear rates, and an exponent around 1 for large shear rates \cite{durian,liu,cloitre}.\\
\indent At the mesoscopic scale, the jamming transition means that the motions of the particles under flow become more and more collective as it is approached, leading to very complex trajectories and rearrangements of the particles.\\

\noindent Actually, this jamming situation has two generic ingredients :
\begin{itemize}
\item{at rest, above a critical concentration, the system is jammed ;}
\item{when a flow is applied to the system, it modifies deeply its dynamic.}
\end{itemize}
 
\indent The effect of flow on a system near jamming transition may be described by a phase diagram shear rate {\it vs} concentration. At zero shear rate, random close packing concentration, $\phi_c$, corresponds to a critical point. Below $\phi_c$, any infinitely small applied stress induces rearrangements and flow of the particules, whereas above $\phi_c$, a finite non-vanishing stress must be applied in order to make the system flow. This means that there is a half-line of degeneracy of the stress --- for vanishing shear rate, above the critical concentration ---, with a critical point at its extremity. But as far as the shear rate is different from zero, the stress always reaches a unique value in a stationary flow, whatever the concentration is. In other words, the shear kills the jammed state, because the shear rate renews constantly the structure of the system\cite{sollich,sollichlong}. This description remains valid as long as the stress remains homogeneous in the sample (no fractures for instance). This phase diagram is very similar to the magnetic field {\it vs} temperature phase diagram of the Ising model. The shear rate $\dot{\gamma}$ plays the role of the magnetic field, the stress $\sigma$ that of the magnetization, and the concentration $\phi$ stands for the temperature. However, Soft Glassy Materials are not at thermodynamic equilibrium, and the relations between $\dot{\gamma}$, $\sigma$ and $\phi$ do not derive from a free energy, but from a dynamical equation. Here, we introduce such an equation, based on a mean-field analysis of a mode-coupling theory.\\

\indent An attempt of this type of approach was recently proposed by P. Sollich, M.E. Cates and us \cite{sollich}. The jamming transition was described using a glass model introduced by J.-Ph. Bouchaud\cite{bouchaud}. One of the results of the model was the scaling law relating the stress with the shear rate, with an exponent depending on the distance to the jamming transition. The control parameter was an effective temperature which attempted to describe the mechanical noise. It was introduced phenomenologically, and was not related to the macroscopic shear rate, whereas shear must induce changes of configurations in the sample, and thus affect its  effective temperature.\\
\indent In this letter, we present a model which describes more directly the interactions between the particles trajectories. This model thus belongs to the mode-coupling class of models\cite{gotze}.\\
\indent We first divide the sample into blocks carrying a stress $\sigma$ depending both on time $t$  and on their position $\mathbf{r}$. The evolution of the blocks follows simple rules :
\begin{enumerate}
\item When submitted to a shear rate $\dot{\gamma}$, the stresses of all the blocks increase during time interval $dt$ by the quantity $G_0\dot{\gamma}dt$.\\
\item If the absolute value of the stress of a given block is larger than a critical value $\sigma_c$, it is set to zero after a time $\tau$. This processus is very similar to rearrangements described by Princen for foams\cite{princen}. However, the fact that the stress is set to zero after the rearrangement, or randomly distributed around zero is clearly not critical.\\
\item If the stress on $\mathbf{r'}$ is set to zero, stress field is modified in the whole sample. We will describe this interaction with a mean field approximation, by adding a random value $\Delta$ to the stresses of all the other blocks .
\end{enumerate}
\indent Let us discuss the last point. Actually, if the block in $\mathbf{r'}$ relaxes from $\sigma$ to zero, the stress at point $\mathbf{r}$ varies of the quantity ${\cal G}(\mathbf{r'}-\mathbf{r})\sigma (\mathbf{r})$ where $\cal G$ is the stress/stress elastic propagator. $\cal G$ is well known in an homogeneous system and it appears that its mean value for the $4 \pi$ steradians is equal to zero. This implies that the random value $\Delta$ must be chosen with a mean value equal to zero. In an homogeneous elastic medium, the stress propagator is short range --- it decreases as $1/r^3$ ---, and very anisotropic, leading to step by step propagation of the stress along preferential directions, and finally fracturation phenomena, as described in earthquakes models \cite{obu}.\\
\indent But, systems we are interested in are not at all homogeneous, so the stress propagates along easy paths defined by successive contacts, and stress thus propagates over a long range. Moreover, these stress paths are very fragile and renew completely after very small variations of the strain, so mean field approach may be a good way to describe these systems.\\
\indent Finally, we can write the evolution for the probability $P(\sigma,t)$ of finding a stress $\sigma$ in a block, at time $t$ :

\begin{equation}
\partial_t P(\sigma,t)=-G_0\dot{\gamma} \partial_{\sigma}P(\sigma,t) 
	+ D\partial^2 _{\sigma^2} P(\sigma,t) \\
	-\frac{H(\vert\sigma\vert-\sigma_c)}{\tau}P(\sigma,t) +
	\frac{1}{\tau}\int_{\vert\sigma\vert >\sigma_c}P(\sigma,t) d\sigma\delta(\sigma)
\label{evolution}
\end{equation}

\noindent where $H(x)=1$ for $x>0$, and $0$ otherwise, and $\delta$ is the Dirac function. \\
The first term corresponds to the first rule, the second one to the noise of the third rule, and the last two terms express the failure for stresses above $\sigma_c$ with a rate $1/\tau$ and the reset of the stresses to zero after the failure. 
The third condition writes :
\begin{equation}
D=\alpha \frac{1}{\tau}\int_{\vert{\sigma}\vert>\sigma_c}P(\sigma,t) d\sigma
\label{couplage}
\end{equation}
where $D$ is the amplitude of the noise, and is then proportional to the density of blocks that rearranged during time $\tau$, by a proportional factor $\alpha$, that depends {\it a priori} on the microscopic properties of the material. It could represent something like a mechanical fragility.\\
\indent	Let us first describe the solution of this model in the absence of shear. We get a naive description of a mode-coupling theory of glass transition \cite{gotze}. The model exhibits a dynamic transition, between a frozen system, and a liquid state in which the stress is self-sustained, through a diffusive term, by over-stressed regions. This description of a jamming transition is very similar to the one due to Ivanov {\it et al.} \cite{ivanov}.\\

\indent For $\dot{\gamma}=0$, one easily finds that ~(\ref{evolution}) and ~(\ref{couplage}) leads to :

\begin{equation}
D=\frac{\alpha D}{1+2\sqrt{D}+2D}
\end{equation}
\noindent For $\alpha<\frac{1}{2}=\alpha_c$, the only solution is $D=0$. Solutions correspond to distributions of stable states of the local stress, {\it i.e.} distributions in which $P$ is equal to zero for $\vert{\sigma}\vert>\sigma_c$ . This is quite obvious : all the terms in ~(\ref{evolution}) vanish in that case.\\
\indent On the opposite, for $\alpha>\alpha_c$ in addition to the previous solution, it exists a stationary solution with a non-zero value of $D$. The system can continuously rearrange as a function of time. It is easy to see that $D$ scales simply as $(\alpha-\alpha_c)^2$ above $\alpha_c$ .\\
\indent	Let us now describe the solution in the presence of a stationary shear rate, $\dot{\gamma} = c^{te}$. In that case, equations (\ref{evolution}) and (\ref{couplage}) have always a single stationary solution (Fig.~\ref{proba}). The derivation is straightforward. We first solve equation (\ref{evolution}) for a given value of $D$, and then write the self-consistent equation for $D$ {\it i.e.} eq.~(\ref{couplage}). Then, we make expansions of this self-consistent equation to get the scaling laws given further. From the distribution probability $P$, we compute the macroscopic stress, $\sigma_{macro}$, {\it via} the mean field assumption : the macroscopic stress is approximated by the average value of the local stresses.\\
\indent Whatever the value of $\alpha$, at high shear rates ($\dot{\gamma}  > 1/\tau $), the flow is Newtonian, and the viscosity scales as the life-time $\tau$ of the excited state multiplied by the elastic modulus $G'$. Collective effects are not important in this situation.\\
\indent For small values of the shear rate, the system exhibits a Newtonian regime in the liquid phase, and a yield stress in the jammed phase. The yield stress appears because the solutions of (\ref{evolution}) and (\ref{couplage}) are degenerate for a vanishing value of the shear rate, leading to a discontinuity of the shear stress as a function of the shear rate.\\

\indent	We find the following exponents, in the limit of very small shear rates :
\begin{eqnarray}
\mathrm{for}\ \alpha > \alpha_c  &,& \sigma\build{\sim}_{\dot{\gamma}\rightarrow 0} G\dot{\gamma}\tau (\alpha-\alpha_c)^{-2}\\
\mathrm{for}\ \alpha < \alpha_c  &,& \sigma\build{\sim}_{\dot{\gamma}\rightarrow 0} G (\alpha_c-\alpha)^{1/2}
\end{eqnarray}
For the critical value $\alpha=\alpha_c$ , we find a scaling between the stress and the shear rate:
\begin{equation}
\mathrm{for}\ \alpha=\alpha_c ,\  \sigma \build{\sim}_{\dot{\gamma}\rightarrow 0} \dot{\gamma}^{1/5}.
\end{equation}

Moreover as far as $ \alpha $ is in the vicinity of $\alpha_c$, the stress always scales like $\dot{\gamma}^{1/5}$ in the crossover regime between the high shear rate and low shear rate limits. (Fig.~\ref{sigma},~\ref{phase}).\\ 
\indent These many regimes mimic the recent experimental results of M. Cloitre on microgel beads around the close packing of the beads, and that of Mason  {\it et al.} on concentrated emulsions \cite{mason}. It describes also the results of D. Durian's simulations of foams above the close packing, and some of his recent experiments.
The value $1/5$ of the exponent may seem odd ; it's not so far from the experimental results of Cloitre ($0.1$), and from simulations of foams using the Durian model by A.Liu ($0.16$)\cite{liuprive}, as well as the experimental results on foams by Durian ($.15$)\cite{durianprive}. \\
\indent	The main other rheological properties of concentrated systems may be studied thanks to the response to a periodic flow, $\gamma(t)=\gamma_0\cos{\omega t}$, expressed in terms of the elastic and loss modulus. In order to calculate the elastic and the loss modulus for a non-vanishing amplitude of the strain, we make an expansion of the evolution equation to the second order in $\gamma$. The calculation is analytical, and we only give the final results.\\
\indent First, we obtain that the response is Maxwellian at low frequency, whatever the values of $\alpha$ and $\gamma_0$. In other words, $G'$ and $G''$ scale respectively as $\omega$ and $\omega^2$ when $\omega \rightarrow 0$. \\
\indent At high frequency, the behavior is more complex : whatever the phase, $G'$ is constant, whereas $G''$ exhibits a maximum, $\omega_c$. It decreases as $\omega^{-1/2}$ in the liquid phase, and as $\omega^{-2}$ in the glassy phase. The more striking result lies in the dependance of the characteristic frequency $\omega_c$ with $\alpha$, which depends on the amplitude of the deformation. Indeed, for vanishing amplitudes, $\omega_c=0$ in the glassy phase, and $\omega_c\sim (\alpha-\alpha_c)^2$ in the liquid phase. This is quite obvious, as the high frequency modulus is nearly constant, $\omega_c$ reflects the behavior if the viscosity. But, for non-zero amplitudes, $\omega_c \build{\sim}_{\gamma\rightarrow 0} \vert\gamma\vert$ in the glassy phase, and  $\omega_c \build{\sim}_{\gamma\rightarrow 0} (\alpha-\alpha_c)^2+\gamma^0$ in the liquid phase (Fig.~4).  \\
\indent So, in the glassy phase, a periodic deformation induces a continuous flow of reorganizations playing the role of an effective temperature. This induces a relaxation process with a characteristic time scaling as $\vert\gamma\vert^{-1}$. It thus exists in this phase infinite non-linearities, with an apparent slow relaxation.\\ 
\indent The important point lies in the result that a finite strain amplitude modifies completely the divergence of the typical time of mechanical relaxation. This has to be compared with experimental results showing that pasty systems exhibit very non-linear viscoelastic behavior . Our model leads to an infinitely small linear regime in the jammed state.\\
\indent	So, our model describes in a very naive way a jamming transition, and mimics many of the experimental results on soft glassy materials. We hope that this primitive model opens new gates to further more subtle descriptions of the jamming transition. Questions arise whether there is some universality or not in these systems. The experimental situation is not clear, as most of the mechanical measurements do not deal with the jamming transition itself and its scaling. Moreover, the spatial homogeneity of the flow depends dramatically on the systems : some of them fracture macroscopically in flow, while other do not, for completely unknown reasons , and mechanical properties are very inhomogeneous. More than ever, precise spatial averagings of the mechanical properties are really a challenge in these jamming systems under flow.\\

{\it Acknowledgements :\ \ \ } We are deeply indebted to S.Obukhov for many discussions. We have also benefited from discussions with T. Witten. This research was suported in part by the N.S.F. Grant n$^\circ$ PHY94-07194. F.L. thanks A. Ajdari for its fruitful hospitality in L.P.C.T..

\bibliographystyle{unsrt}
\bibliography{jamflow}

\vspace{1 truecm}

\newpage

\begin{figure}  
\caption{Distribution probability $P(\sigma)$ of the stress, for different values of $\alpha$.}
\vspace{2 truecm}
\label{proba}
\end{figure}

\begin{figure}  
\caption{Stress $\sigma$ {\it vs} shear rate $\dot{\gamma}$ for different values of $\alpha$. }
\vspace{1 truecm}
\label{sigma}
\end{figure}

\begin{figure}
\caption{Dynamical phase diagram : yield stress $\sigma_c$ {\it vs} $\alpha$ for different values of the shear rate $\dot{\gamma}$.}
\vspace{1 truecm}
\label{phase}
\end{figure}

\begin{figure}
\caption{Behavior of the characteristic frequency, $\omega_c$ {\it vs} $\alpha$ at constant peak amplitudes of the shear.}
\vspace{1 truecm}
\label{oma}
\end{figure}

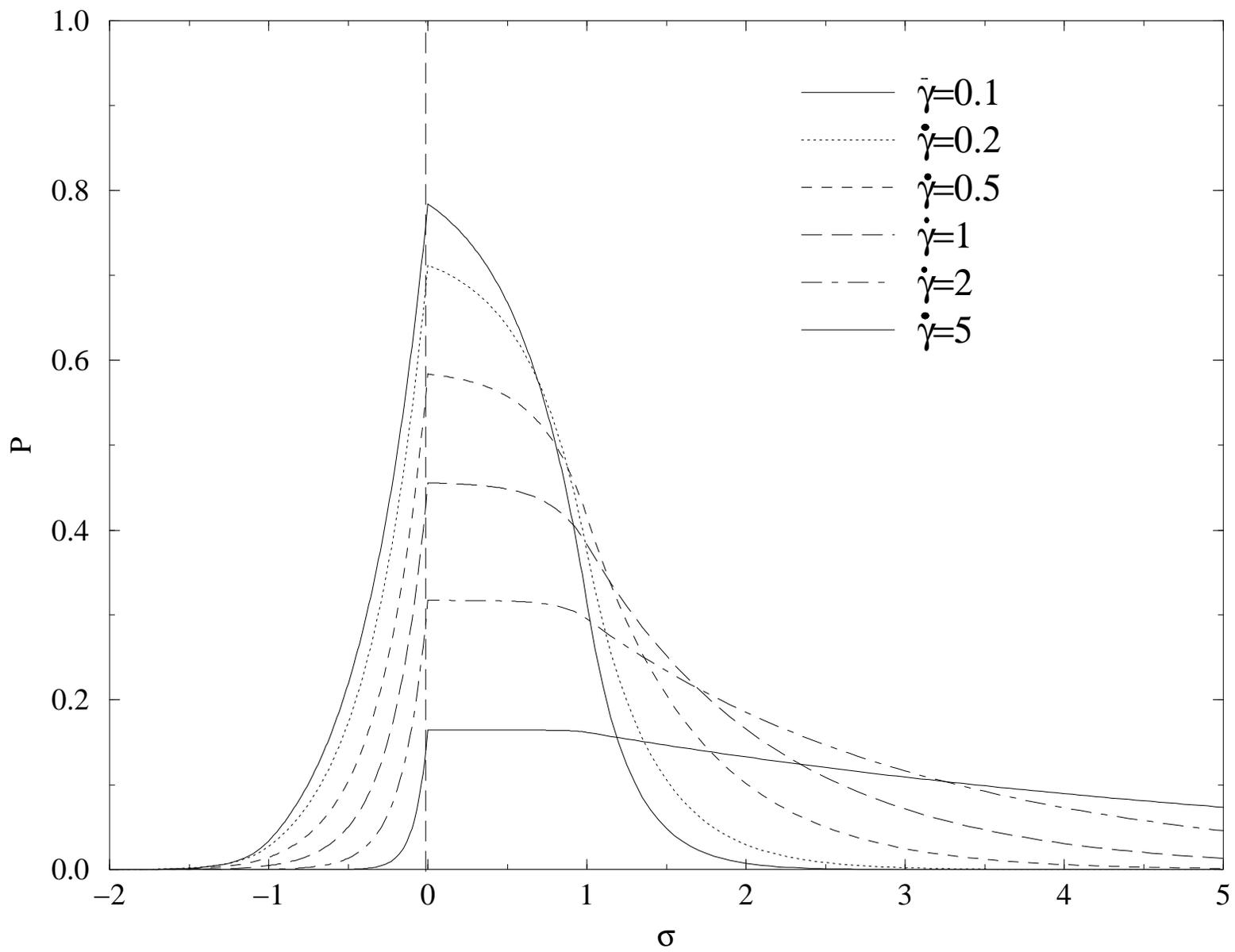
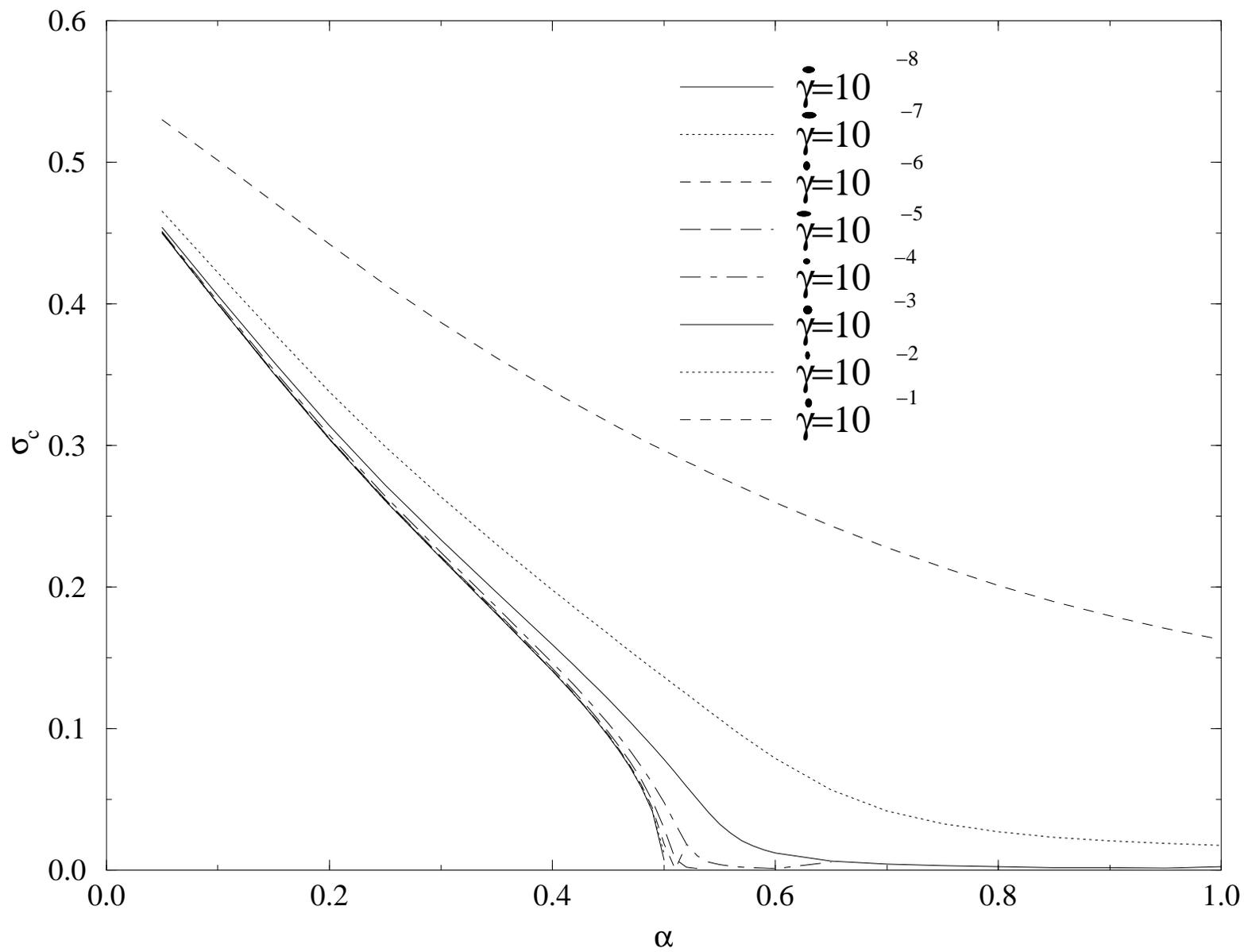
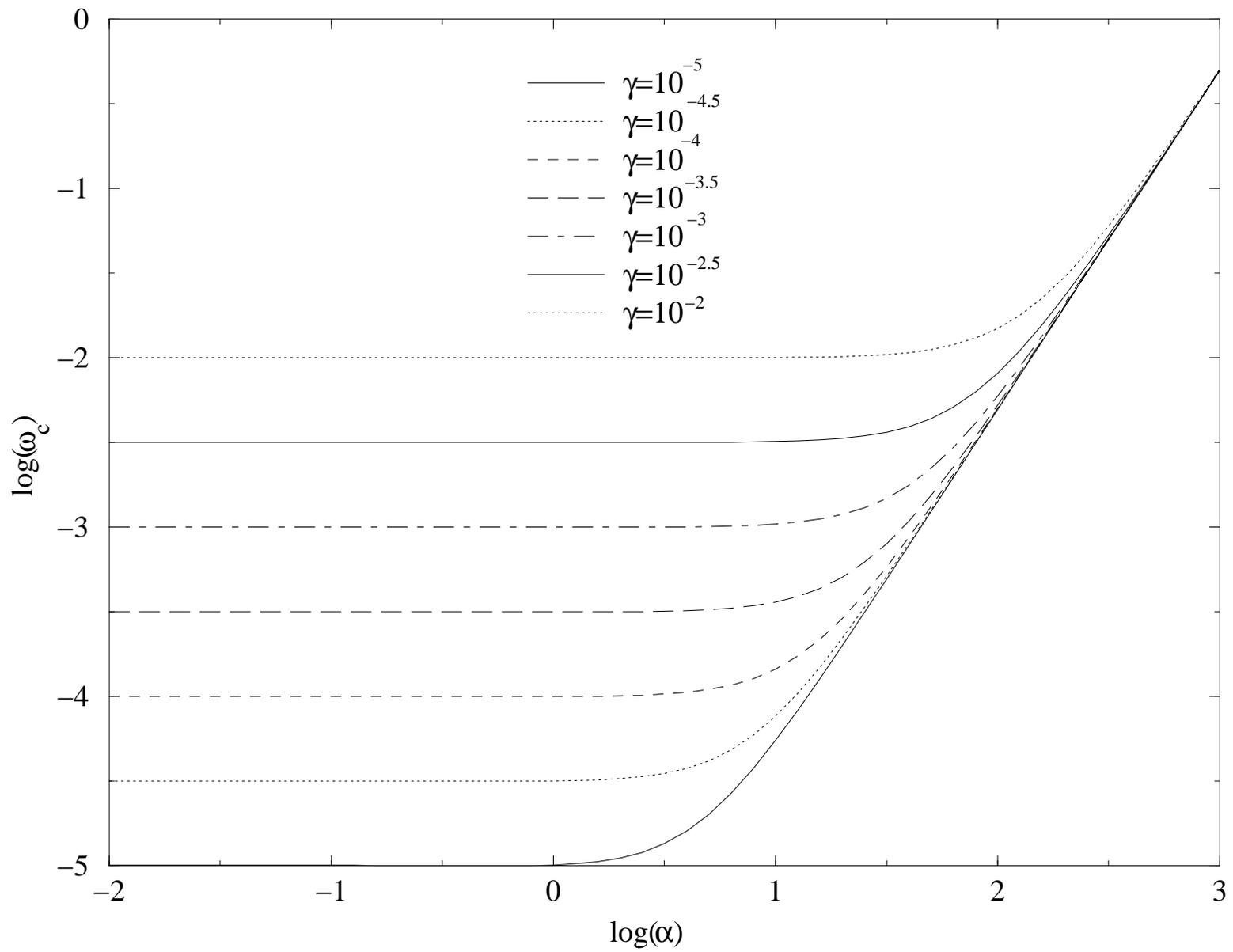
\end{document}